\documentclass[a4paper,11pt]{article}
\usepackage{jinstpub} 
\usepackage{graphicx}
\usepackage{subcaption}

\title{\boldmath UV hybrid photon detector based on GaN photocathodes and Si low gain avalanche diode}

\author[a]{Mohamed Boukhicha}
\author[a]{, Thomas Y. Tsang }
\author[a]{, Gabriele Giacomini }
\author[b]{, Amir M. Dabiran }
\author[a]{, Luca Cultrera}
\affiliation[a]{Instrumentation Division, Brookhaven National Laboratory, Upton, NY 11973}
\affiliation[b]{Qrona Technologies, Eden Prairie, MN 55344}
\emailAdd{mboukhicha@bnl.gov}

\abstract{Photon detectors featuring single-photon sensitivity play a crucial role in various scientific domains, including high-energy physics, astronomy, and quantum optics. Fast response time, high quantum efficiency, and minimal dark counts are the characteristics that render them ideal candidates for detecting individual photons with exceptional signal-to-noise ratios, at frequencies in the the range of hundreds of MHz.
Here, we report on our first design and operational results on a Hybrid Photon Detector (HPD) that combines the high quantum efficiency of a Gallium Nitride (GaN) photocathode and the low noise characteristics of a Si-based Low-Gain Avalanche Diode (LGAD). This hybrid detection scheme has the potential to reach single-photon detection sensitivity with high quantum efficiency, low noise levels and capable of operating at hundreds of MHz repetition rates.}


\begin{document}
\maketitle
\flushbottom

\section{Introduction}
\label{sec:intro}
Single photon detectors are of interest for many applications ~\cite{Leutz2003,HPD-3-5}. 
Although various variances of Hybrid Photon Detector (HPD) have been explored ~\cite{compactHPD,HighSpeedHPD,Abalone,Ablone2}, the ones offering simultaneously large sensitive area and position sensitive detection cannot yet compete with the performances of proximity focused photomultiplier tubes coupled with microchannel plates (PMT-MCP) detectors ~\cite{JORAM1999407, LYASHENKO2020162834}.  Lowering the cost per unit of large area fast position-sensitive photon detectors while preserving the capabilities of single photon sensitivity, low noise, high spatial and picosecond scale time resolution is one of the challenges that need to be overcome to develop the sensors that can enable next generation experiment requiring detectors with tens of square meters of active area.   
In this prototype, the UV photons penetrate the fused silica vacuum window and illuminate a Negative Electron Affinity (NEA) activated GaN photocathode, which is negatively biased to a few kV, thus releasing photoelectrons. These photoelectrons are then accelerated to energies of a few keV and directed towards a Low-Gain Avalanche Diode (LGAD). Upon reaching the LGAD, the electrons undergo an electron bombardment process, depositing their kinetic energy and producing, on average, one electron-hole pair for every 3.6 eV deposited in the silicon substrate. This avalanche multiplication process results in the generation of additional electric charge within the LGAD, allowing for efficient detection of the incident electrons. 
\begin{figure*}[h!]
\centering
\includegraphics[width=0.7\linewidth]{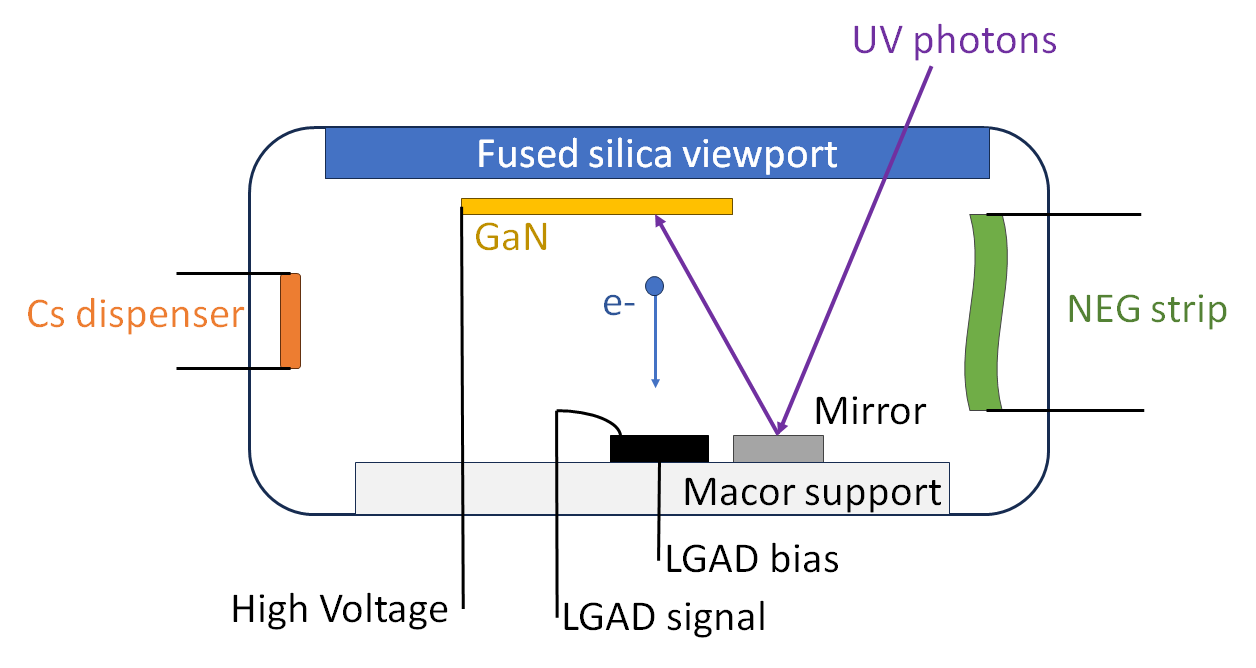}
\caption{HPD schematics, illustrating the working principle of the experiment.}
\label{fig:HPD 3D CAD model}
\end{figure*}
The total number of electron-holes pairs generated in this step will scale linearly with the accelerating voltage applied between the photocathode and the LGAD. The secondary electrons generated inside the LGAD structure will then drift toward the p-n junction 
where an intense internal electric field triggers an avalanche of electron-hole pairs measured by the electronic output. 
The HPD prototype we describe in this paper was operated in continuous wave (cw) and pulsed mode operation using UV photons generated by an LED. We used the cw operating mode to perform alignment of the light source and to provide preliminary measurements of the photon detection response; while we employed the pulsed operating mode to examine the potential for the keV single photoelectron detection in the LGAD.

The expected total gain of the HPD is the product of the number of electron-hole pairs generated in the primary electron bombardment of the LGAD active area and the avalanche gain in the LGAD. By combining the typical electron multiplication from impact ionization ($\sim 10^3$) with the LGAD amplification ($\sim 10$) it is possible to achieve charge gain exceeding $10^4$. This hybrid configuration aim at taking advantage of fast response times and low noise, typical of LGAD devices, and expanding their application to the detection of single-photon in the UV region of the spectrum. The strategic use of HPD in the proximity focused geometrical configuration 
minimize the transit time spread of photoelectrons allowing to fully leverage the LGAD exceptional temporal resolution and capability of operating at hundreds of MHz repetition rates ~\cite{Giacomini2021,GALLOWAY20195}. Also typical low dark counts from both GaN photocathodes and LGAD  will make this device operable already at room temperature ~\cite{Siegmund2008}. The proximity focusing geometry also allows an easy transition to finely segmented LGADs to achieve high spatial resolution. Using the state-of-the-art high quantum efficiency of GaN and ternary alloys AlGaN and InGaN, the spectral sensitivity of the HPD can be tuned by proper tailoring of the photocathode bandgap. 
Unlike other HPD or MCP-PMT devices that rely on the use of alkali antimonides or telluride photocathodes for the generation of the primary electron, the assembly of our device, which can be considered a \textit{"vacuum tube"}, does not require Ultra High Vacuum (UHV) environment. Other HPD and MCP-PMT devices operated with contamination sensitive alkali antimonides or tellurides photocathodes cannot be exposed to atmospheric pressure, otherwise resulting in the irreversible damage of the photoemission properties of the photocathodes.
In our approach, the  GaN photocathode is activated to the condition of Negative Electron Affinity by exposing it to Cs vapors only after the device has been assembled and the vacuum condition established in the HPD vessel.
This modular approach allows the development of different components (mainly the photocathode and the LGAD) separately from their final packaging.
Motivated by the application in consumer opto-electronics (Light Emitting Diodes) and high power electronics (High Electron Mobility  Transistors), the technology for the synthesis of GaN and other III-nitride semiconductors has become very reliable due to large investment from private sector. 
Extending the use of III-nitride semiconductors to other applications such as photon detection can leverage the already advanced state-of-the-art industrial production of these materials.

\section{HPD design:}
The goal of this work was to demonstrate a proof-of-principle robust HPD module whose assembly process does not require to be carried in an ultra-high vacuum environment, allowing to be easily scaled for mass production.
To reach this objective, we envisioned an assembly process for handling of photocathode and LGAD in a controlled atmosphere at ambient pressure conditions rather than in an UHV environment. The choice of GaN photocathode material was dictated by the objective of avoiding assembly in UHV. We believe that with this design, the photodetector can reach performances on par with other HPDs while reducing overall costs when compared to other photon detectors that use alkali antimonide photocathodes, which require complex and time-consuming processing in UHV environment.

\begin{figure*}[h!]
\centering
\includegraphics[width=0.7\linewidth]{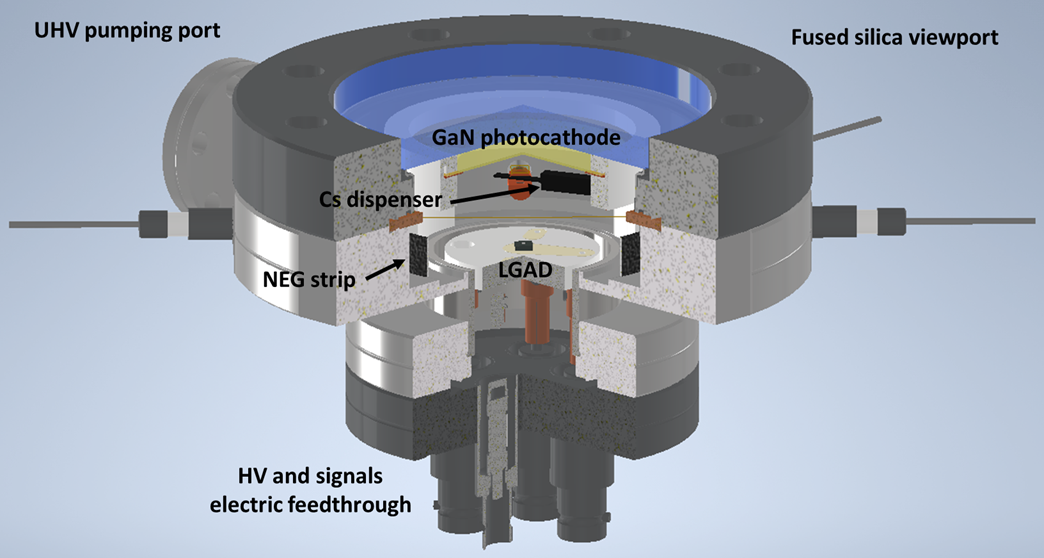}
\caption{HPD 3D CAD design }
\label{fig:V-design}
\end{figure*}
 
 Figure~\ref{fig:V-design} depicts a three-dimensional CAD model of the HPD prototype, its working principle was described above (Figure \ref{fig:HPD 3D CAD model}).
 Electric signals are read-out by a charge sensitive amplifier followed by a pulse shaping amplifier, and the charge signal is displayed on a fast digital oscilloscope.

NEA activated GaN photocathodes are capable of efficiently converting incident UV photons into photoelectrons as demonstrated by quantum efficiency up to 70\% when illuminated with deep UV photons\cite{Uchiyama2005}. On the other hand, due to its intrinsic wide bandgap (~3.5 eV), GaN is sensitive to wavelengths shorter than ~350 nm \cite{Sieg2006DevelopmentOG}. With the use of ternary alloy based on InGaN the bandgap may be tuned from the UV to the IR extending the spectral response of the HPD. 
The use of GaN as the photocathode  offers several advantages such as high quantum efficiency, low dark counts, and robustness under various environments due to its wide bandgap\cite{Sieg2008GalliumNP,Sieg2006DevelopmentOG, Siegmund2008}. Therefore, GaN photocathodes are well-suited for applications where single UV photon sensitivity is required. 

\begin{figure}
\centering
\includegraphics[width=0.5\textwidth]{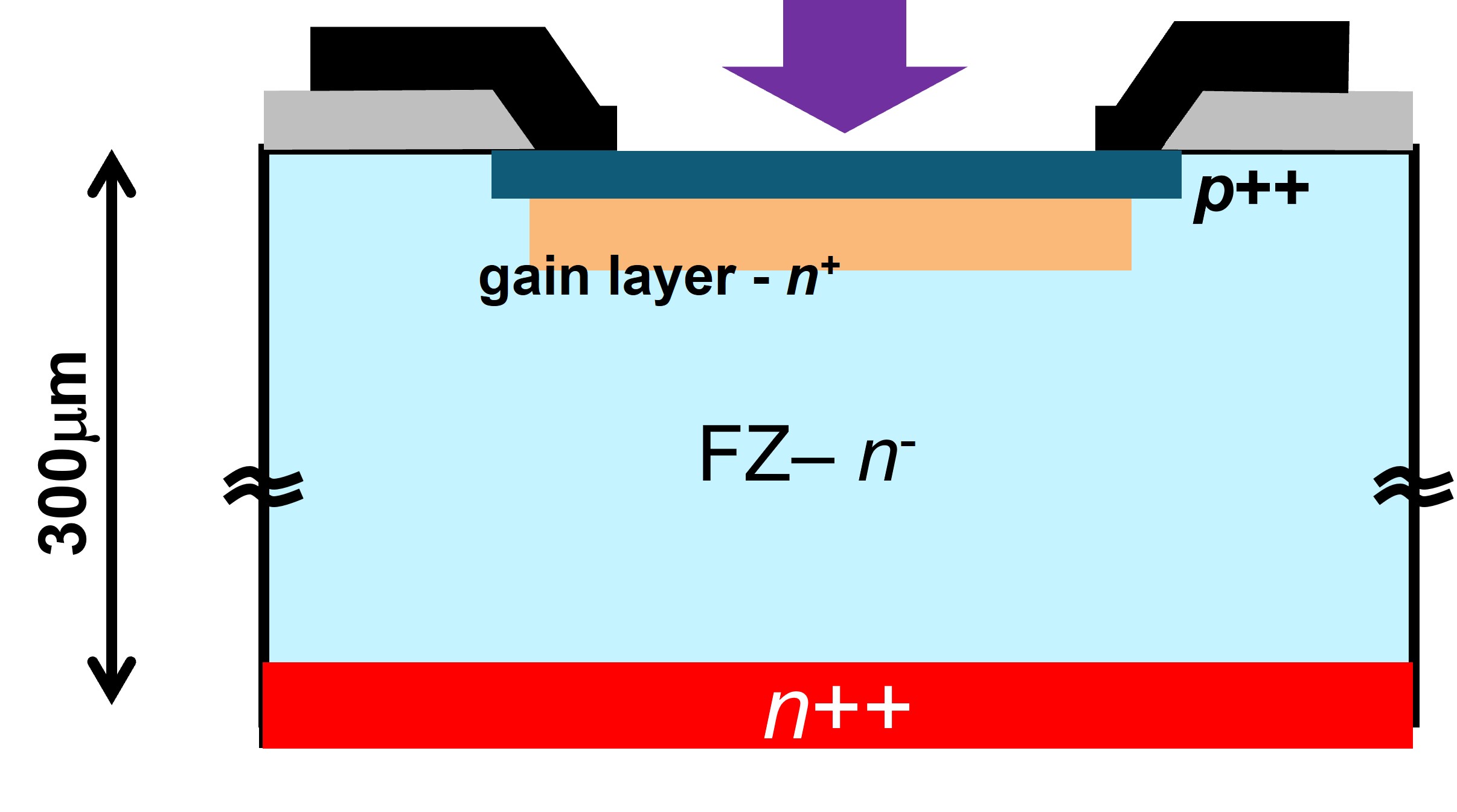}
\caption{Sketch of the section of the LGAD for low-penetrating particles used in this study.}
\label{fig:LGAD scheme}
        
\end{figure}

LGAD is a class of silicon sensors originally developed for the fast detection of Minimum Ionizing Particles (MIP) \cite{Giacomini2019}. LGADs leverage their thin silicon substrate and built-in amplification to achieve a higher sensitivity than a standard silicon diode. In this sense, the LGAD operation and structure are analogous to that of a standard Avalanche Photo Diodes (APD) but with a lower gain ($\sim10$ vs. 100-1000), and therefore with a significantly lower noise level, thus improving the signal-to-noise ratio of the detector. For this project, we used a readily available small-area LGAD, whose doping profiles are of the opposite polarity with respect to those of a regular LGAD, see Fig.\ref{fig:LGAD scheme}. Such structure allows the multiplication of low penetrating particles, electrons in our case. The thickness of the used LGAD is 300 $\mu$m.
The amplification mechanism is obtained by introducing an $n^{+}$ layer between the $p^{++}$ and n-bulk layers. Because of the $n^{+}$ implant, the electric field across the $n^{+}/p^{++}$ junction is high enough to trigger an avalanche process for the signal  electrons drifting towards the backside of the device. Our LGAD sample has an active area of 0.42 mm$^{2}$ (Figure~\ref{fig:LGAD scheme}). A guard ring structure surrounding the central pad serve to shield the active area from external currents or signals.  Details of the in-house design, fabrication, and electrical characterization of this and other LGADs can be found in previous works \cite{Giacomini2019,Giacomini2021}. 

\begin{figure}[htbp]
    \centering
    \begin{subfigure}[b]{0.4\textwidth}
        \includegraphics[width=\textwidth]{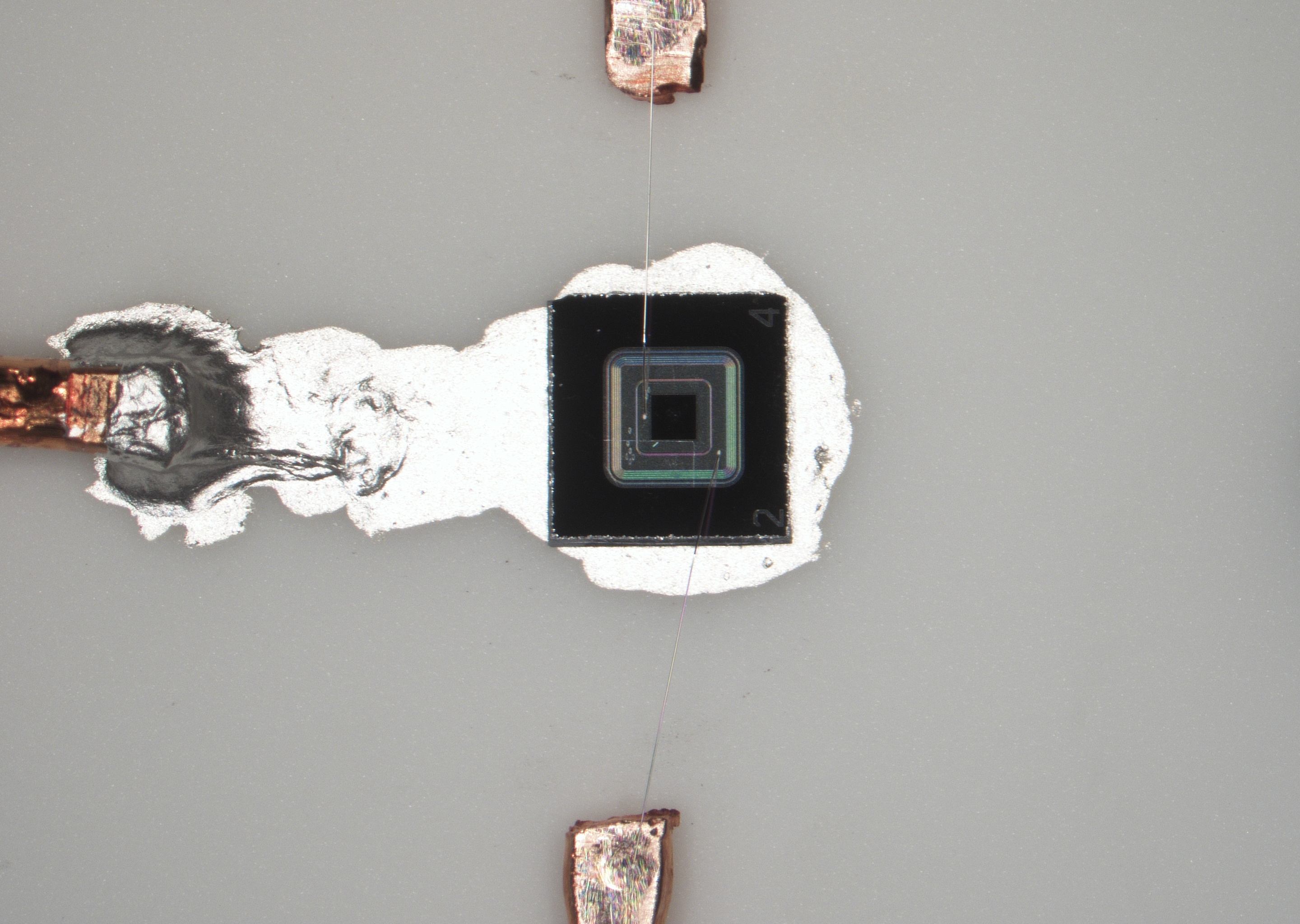}
        \caption{LGAD glued on the Macor plate with silver paint. Wire bonding to guard ring and collector are also visible.}
        \label{fig:LGADchip}
    \end{subfigure}
    \qquad
    \begin{subfigure}[b]{0.4\textwidth}
        \includegraphics[width=\textwidth]{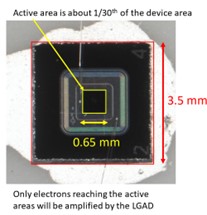}
        \caption{Picture of the LGAD as connected to the ceramic base plate inside the HPD.}
        \label{fig:design}
    \end{subfigure}
    \caption{Pictures of the LGAD as connected to the ceramic base plate inside the HPD.}
    \label{fig:LGAD}
\end{figure}

The HPD assembly can operate with semi-transparent photocathodes in transmission modes or, as in the presented results, with opaque photocathodes operated in reflection mode. In this latter case, a small UV enhanced aluminum optical mirror is placed inside the detector to reflect the photons towards the active reflection-mode photocathode surface.
A non-evaporative getter (NEG) strip is placed in the vacuum vessel to maintain an ultra high vacuum environment.
A metal chromate cesium dispenser is placed inside the vessel to provide alkali metal for the initial and periodic NEA activation of the photocathode, if needed.
The vacuum enclosure for our HPD detector prototype is based on commercially available off-the-shelf UHV components. The UHV vessel is built using 4.5" Conflat hardware: a fused silica viewport, a double-sided flange as main body and a reducer nipple 4.5" to 2.75". The double-sided body flange has custom ports for the electric feedthroughs required for NEG strip, Cs dispenser and vacuum pumping port.
The 2.75" flange at the bottom has the BNC connectors used to provide LGADs and photocathode bias and collect signals.
The LGAD device body is mechanically connected to a ceramic (Macor) plate using silver paint which also is used to provide electric contact to bias the diode while wire bonding is used to connect the guard ring and collector to the copper wires and the BNC feedthroughs (see figure \ref{fig:LGAD}).

The GaN photocathode is held by a ceramic support structure which keeps it at 20 mm above the LGAD surface. Electric contact between the copper shim in the support structure and surface of the GaN photocathode is made using silver paint.
The copper shim is connected to a SHV-BNC feedthrough which carries the high voltage bias for the acceleration of the photoelectrons. The GaN photocathode is operated in reflection mode. To illuminate the photocathode the light from the UV light source is sent over a mirror placed near the LGAD and glued to the same Macor baseplate using silver paint. The photons are then reflected  by the mirror onto the GaN surface.
Altough the prototype was realized using commercially available UHV parts to minimize customization, a fully developed HPD will have a significant higher ratio between active and overall area. 
An UHV all-metal angle valve is connected to the single vacuum pumping port of the device, allowing the evacuation of the internal volume after the device is assembled and sealed. After a seven days long 150 C bake out of the HPD device, the NEG module is fully activated and the valve at the pumping port is closed.
In our first test, a miniature ionization gauge was installed with a vacuum tee on the pumping port to measure the effectiveness of the NEG strip in holding the vacuum level after the valve was sealed. The vacuum level measured by the gauge after the HPD sealing was $1\times10^{-9}$mTorr. We expect the base pressure on the HPD volume to be lower than this value because the vacuum gauge volume is pumped via a low conductance (about 0.1 l/s) tube. 
 on affinity by lowering the vacuum level of the GaN photocathode resulting in an efficient electron emission \cite{Qian2009ActivationAE}. Thus, GaN is capable of efficiently converting incident UV photons into photoelectrons up to 70\% high quantum efficiency \cite{Uchiyama2005}. 
 tuned from the UV to the IR extending the spectral response of the HPD. 

\subsection{GaN activation and response}

Prior to characterizing the performance of the HPD detector, the GaN photocathode was submitted to a cesiation process to bring the surface to the NEA condition and to achieve a high quantum efficiency. The Cs dispenser was degassed by passing $\sim$3A of dc electric current through it for a few minutes toward the end of the UHV bake-out procedure, before sealing the angle valve, 
to remove residual contaminants such as water and hydrocarbon physisorbed by the dispenser boat and activate the getter materials.  
Thereafter, cesiation of the GaN photocathode is achieved by increasing the bias current to $\sim$4.5A. At this current the Cs dispenser release high purity alkali metals that deposited a few atomic layers of Cs on the surface of GaN lowering its workfunction and enabling the formation of the NEA condition.
By illuminating the photocathode with a fixed intensity of UV 270 nm photons, the photocurrent of the GaN is continuously monitored during the entire cesiation period to monitor the formation of the NEA activation layer and estimate the achieved quantum efficiency. 

\section{Experimental results}

Due to unexpected thermal drift the Cs dispenser was compromised during the initial degassing process.
Nevertheless, upon exposure to the Cs vapors the photocurrent increased to the level of 20 nA which we estimated corresponds to a quantum efficiency (QE) of 0.11\%. Although the obtained QE in the HPD is below our expectation and sub-par compared to the results usually obtained in other activation experiments,
we deemed the QE level is sufficient to proceed in characterizing the HPD performances rather than to break the vacuum and restart the whole process which would require additional time and resources. 

\subsection{CW measurements}
To better understand the response of our HPD and to eliminate the possibility of space-charge effect, we first performed cw UV photon illumination measurement by focusing a fiber-coupled (Thorlabs UM200, 600 µm diameter UV solarization resistance fiber) 270 nm UV LED light to a spot diameter of ~5 mm on the GaN photocathode in a dark room environment. 
Background signal for subtraction was estimated with and without UV photons' illumination, with and without bias fields on both LGAD and GaN photocathode, respectively. The cw measurement allowed us to accurately align the UV photon illumination spot onto the photocathode surface. 
Using a Thorlabs optical power meter, we measured $105.3\mu$W  of fiber-coupled UV LED on the GaN photocathode taking into account  the UHV viewport transmittance and the in-vacuum mirror reflectance. At a 3 kV bias on the photocathode we measured a photocurrent of 9 nA while the amplified signal current produced by the LGAD at 50V bias was 285 nA. 
While the Si chip with the LGAD structure is $3.5\times3.5$ mm$^2$ the active area within the guard ring structure where the LGAD  amplification is enabled is only $0.65\times0.65$ mm$^2$ (see figure 3b).
Numerical simulation carried out using SIMION software allowed to confirm our expectation that electrons generated from the photocathode will travel perpendicularly to the emitting photocathode surface due to the proximity focusing geometry. Because of that, only electrons emitted from the photocathode areas aligned to the active LGAD region can be detected. Figure 4 shows the simulated scatter plot with location of impact of photoelectrons generated in a 5 mm diameter area centered above the LGAD. Only about 2\% of the emitted photoelectrons will reach the LGAD active area.
This low collection efficiency is only an apparent limitation of our HPD device. Such a limitation indeed is due only to the specific geometry of the current LGAD test device. LGADs with larger active area, with or without segmentation, can significantly increase the photoelectrons collection efficiency.

\begin{figure}
    \centering
    \includegraphics[width=0.5\textwidth]{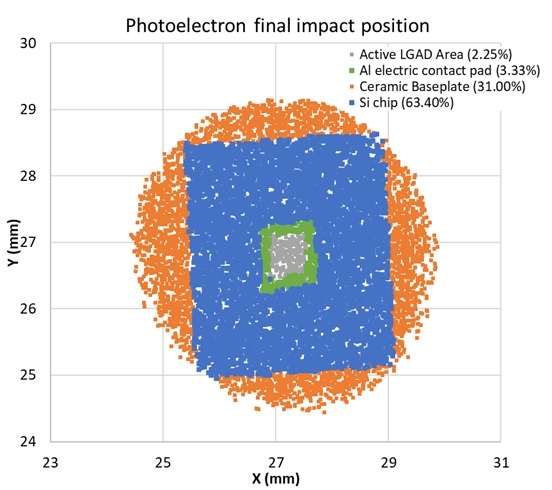}
    \caption{Results of electron trajectory tracking obtained with SIMION. Due to the geometry of the system and experimental conditions it is expected that only $\sim$ 2\% of electrons extracted from the phtocathode will reach the active area of the LGAD.}
    \label{fig:LGAD SIMION}
        
\end{figure}

We define the HPD gain as the measured LGAD current vs the estimated fraction of photocathode current reaching the active LGAD area and then obtain a combined gain of 1580 for a single 3keV photoelectron.
While these cw measurements provide valuable insights into the HPD characteristics under continuous UV photon illumination, discerning the single photon detection characteristics required the implementation of pulsed measurements.

\subsection{Pulsed UV photon measurements:}

\begin{figure*}[h!]
    \centering
    \begin{subfigure}{0.55\linewidth}
        \includegraphics[width=\linewidth]{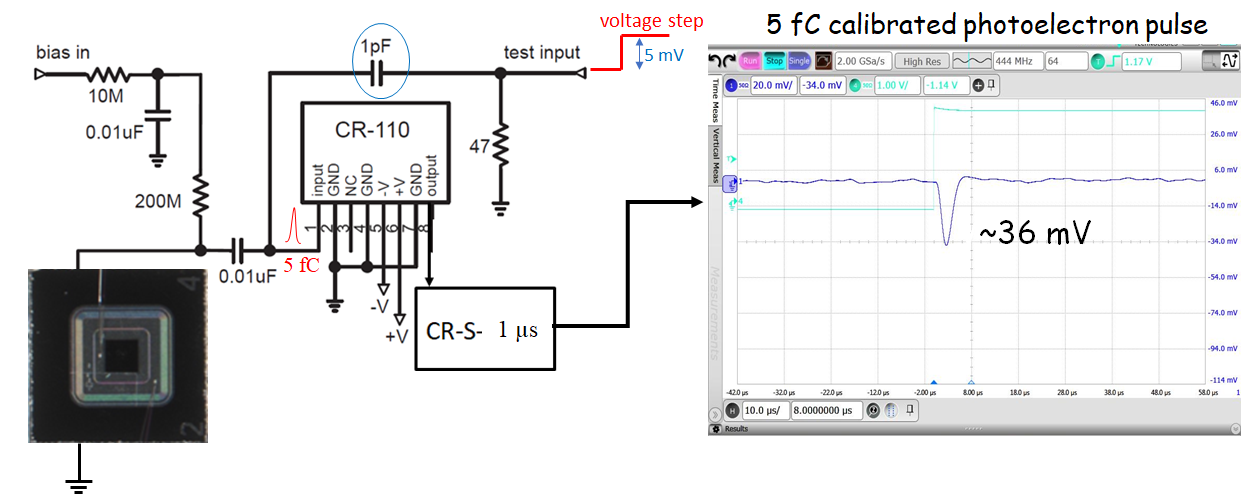}
        \caption{}
        \label{fig:CREMAT}
    \end{subfigure}\hfill
    \begin{subfigure}{0.32\linewidth}
        \includegraphics[width=\linewidth]{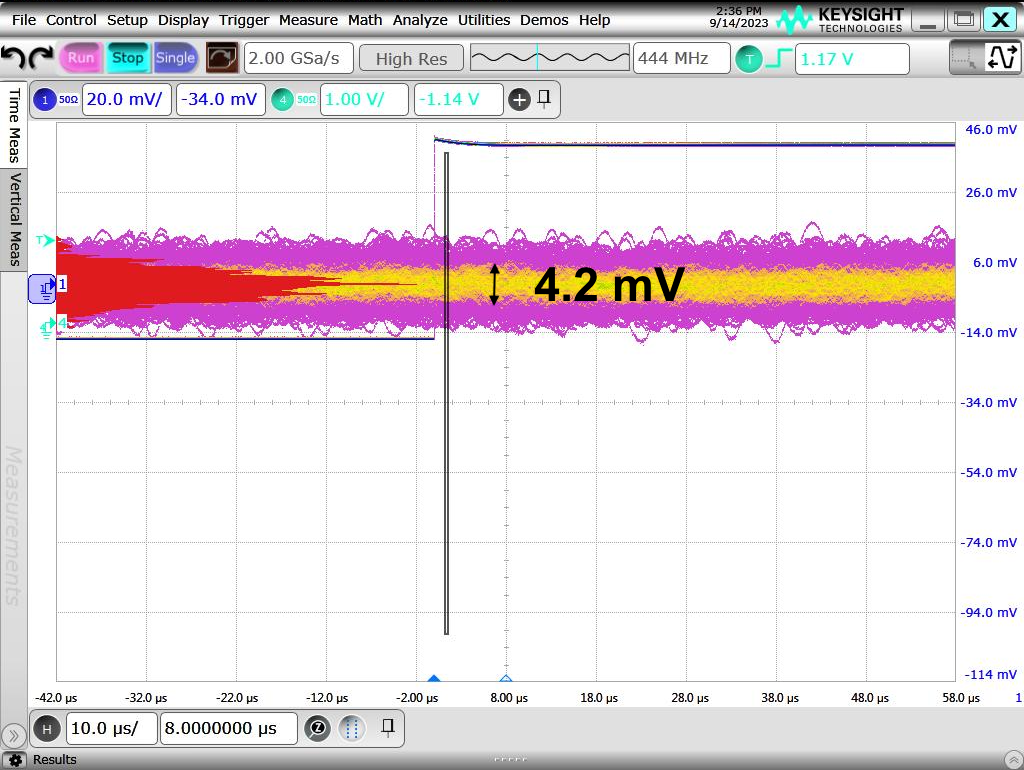}
        \caption{}
        \label{fig:RMS Noise}
    \end{subfigure}
    \caption{ Electronic for charge measurement, (a) charge calibration, (b) system noise}
    \label{fig:design}
\end{figure*}

To collect and process the photoelectron signal, we employ a Cremat CR-110 low-noise charge-sensitive preamplifier and a Cremat CR-S-1µs shaping amplifier, all signal outputs are displayed and measured on an oscilloscope. To calibrate the electronic charge gain, we inject a known 5 fC charge to the system (a 5 mV step voltage onto a calibrated 1pF capacitor) with the LGAD connected and biased to 25 V, resulting in a charge signal of 35.5 mV, see Figure ~\ref{fig:CREMAT}. Therefore, we established our electronic charge calibration to be 0.139 fC/mV (868 electrons/mV).

Given that the electron-hole pair creation energy in silicon is 3.6 eV, a 3 keV photoelectron emitted from the GaN photocathode under 3 kV bias would yield  $\sim833\%$electron-hole pairs. Without any signal amplification, the conversion corresponds to a charge accumulation of $Q = 833 \times (1.602 \times 10^{-19}) = 0.1333 \, \text{fC}$ per incidence of 3 keV electrons.

With the LGAD and GaN photocathode connected and biased but in complete darkness, we measured a system noise of 4.2 mV rms, see Figure~\ref{fig:RMS Noise}, corresponding to a noise charge of 0.59 fC (4.2 mV x 0.139 fC/mV), or  ~3700 r.m.s. electrons. This electronic noise is ~4.4 time higher than the $\sim833\%$ electron-hole pairs generated by a single 3 keV electron. 
With this level of electronic noise, we are able to detect $\sim$5 photoelectrons extracted from GaN with a 3 kV bias.

\begin{figure*}[h!]
    \centering
    \begin{subfigure}{0.48\linewidth}
        \includegraphics[width=\linewidth]{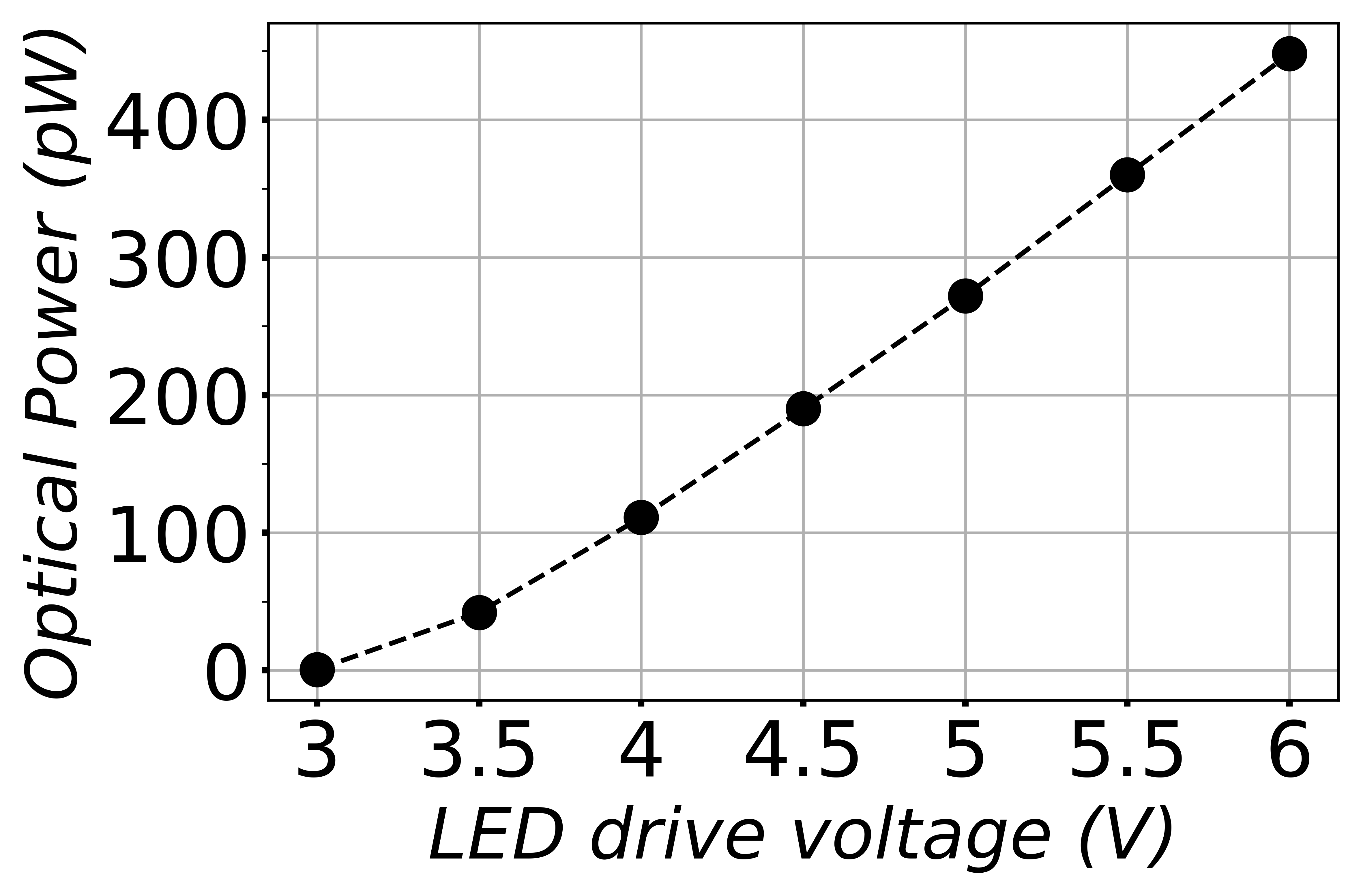}
        \caption{}
        \label{fig:LED_int}
    \end{subfigure}
     \begin{subfigure}{0.46\linewidth}
        \includegraphics[width=\linewidth]{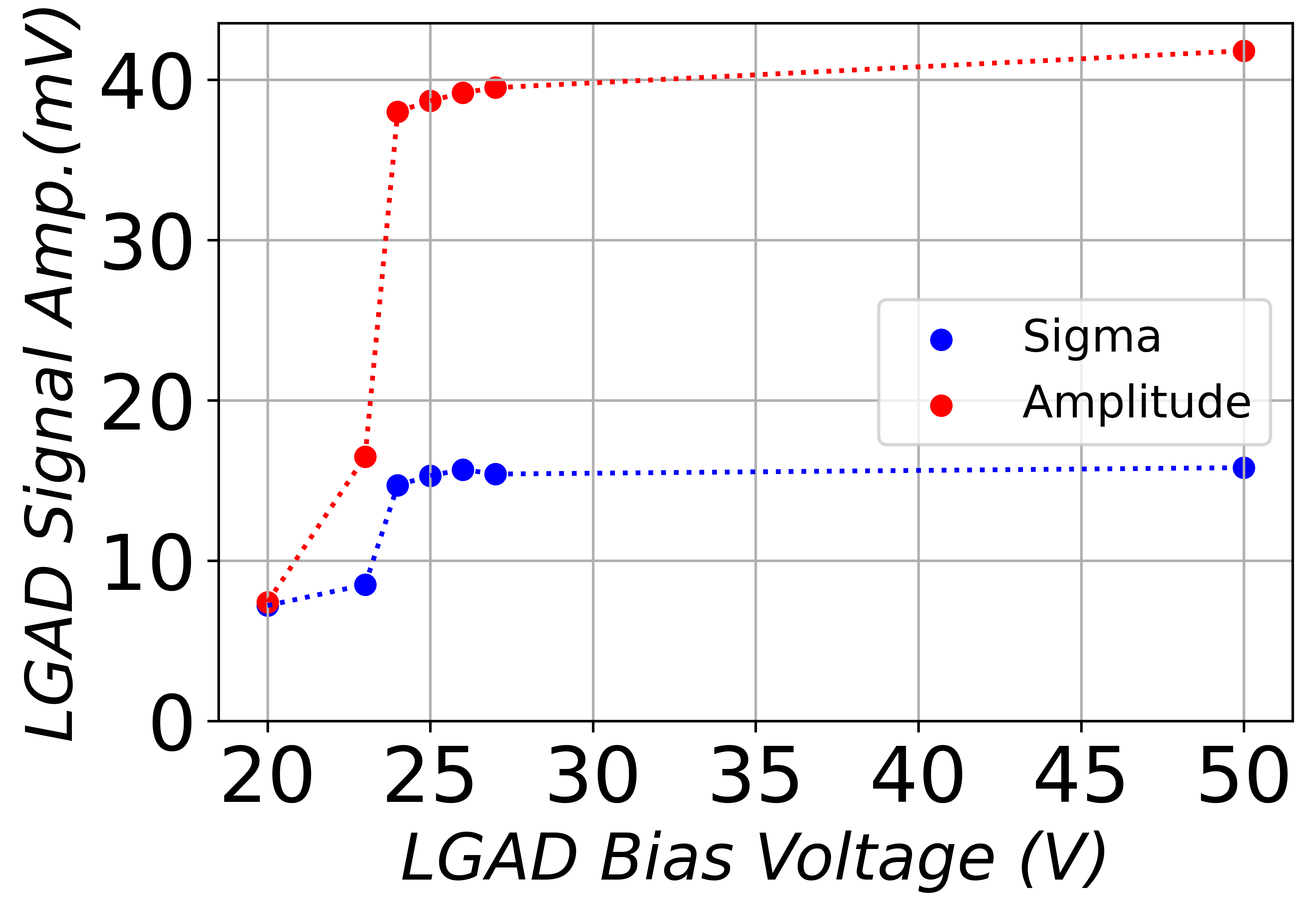}
        \caption{}
        \label{fig:LGAD_pulsed}
    \end{subfigure}\hfill 
    \caption{(a) Linearity of the UV LED output as a function of the drive voltage. (b) charge signal pulse amplitude and width versus LGAD bias. }
    \label{fig:LGAD response}
\end{figure*}

To perform pulsed measurements, the fiber-coupled UV LED is driven with a voltage pulse of fixed amplitude and a temporal width of 20 ns at a repetition rate of 100 Hz. 
We measured the fiber-coupled optical average power versus the drive pulse amplitude in a range from 4V to 6V. A linear optical power output with drive voltage was observed, see Figure ~\ref{fig:LED_int}.

We measured an average optical power of 110 pW when the UV LED is pulsing with a 4V amplitude in the drive signal.
After considering $\sim20\%$ optical loss due to transmission of fused silica viewport and reflectivity of the in-vacuum mirror, we estimated $\sim110pW\times0.80 = 88 pW$ reached the GaN photocathode. With the estimated QE of 0.11\% this corresponds to an average emission rate of $\sim$136000 photoelectron/s from the surface of the GaN photocathode.
Under different bias conditions the response of the LGAD shows a charge gain saturating at the bias voltage of ~26 V, Figure \ref{fig:LGAD_pulsed}, where the signal pulse height distribution is plotted as a function of the LGAD bias voltage for 3 keV photoelectrons. We noted that the pulse width of the charge signal (sigma) remains nearly constant for the detection of 3 keV electrons beyond 24 V bias on the LGAD, see the sigma data in Figure \ref{fig:LGAD_pulsed}, suggesting that the electronic noise and the statistics on the kinetic energy of the photoelectrons dominate the charge signal response of the LGAD. 

\begin{figure*}[h!]
    \centering
    \begin{subfigure}{0.48\linewidth}
        \includegraphics[width=\linewidth]{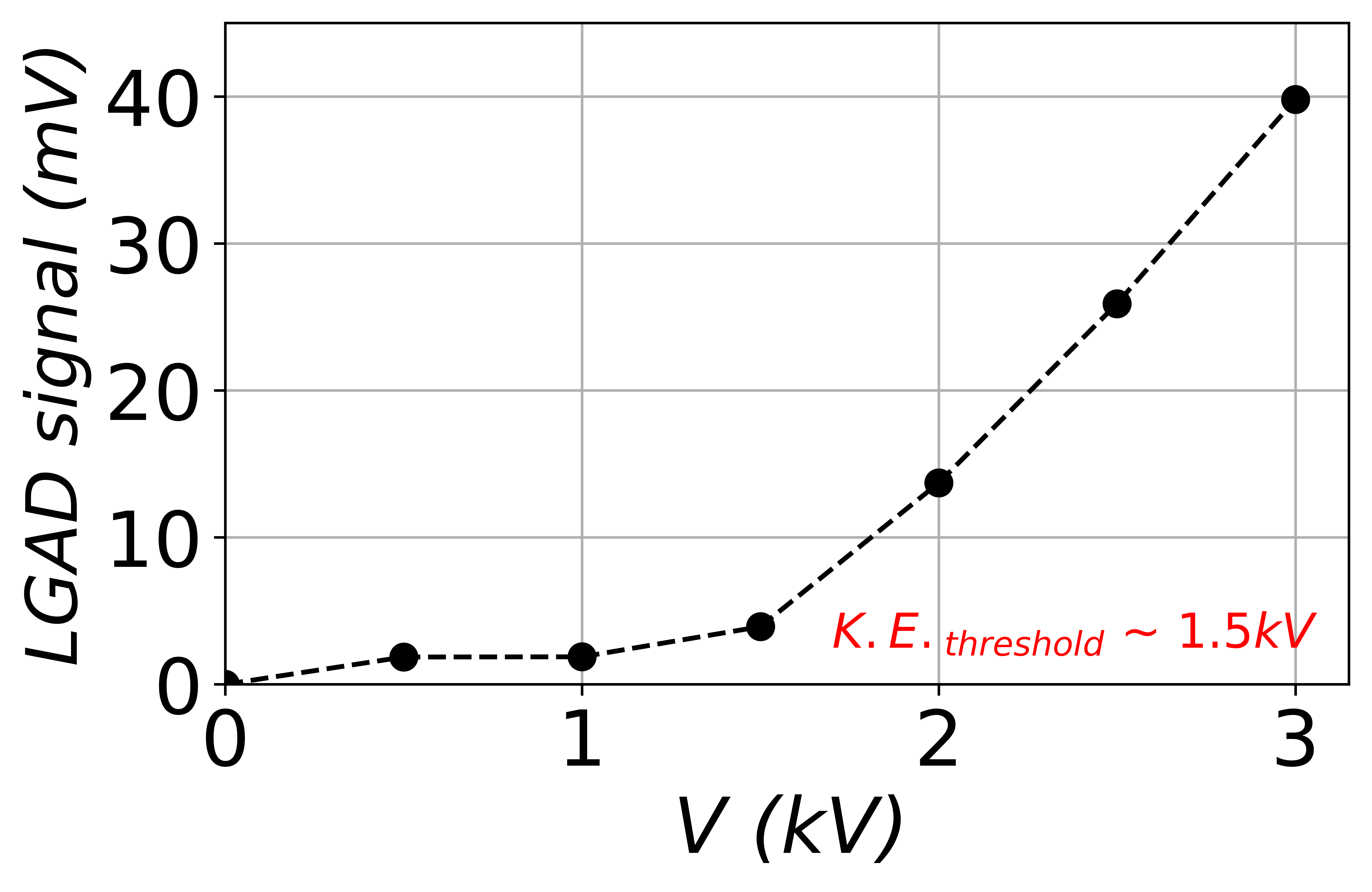}
        \caption{}
        \label{fig:LGAD_HV}
    \end{subfigure}\hfill
    \begin{subfigure}{0.48\linewidth}
        \includegraphics[width=\linewidth]{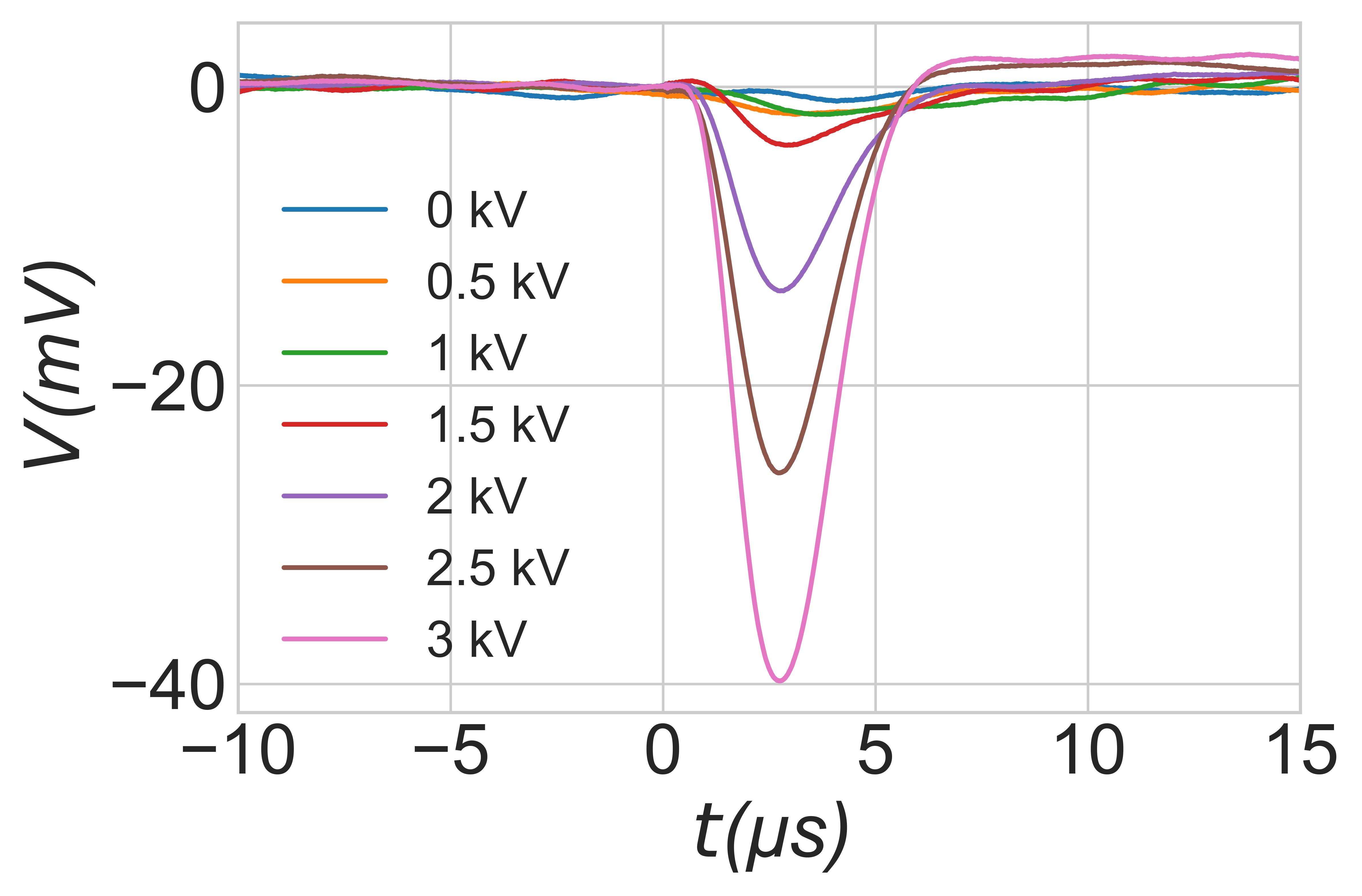}
        \caption{}
        \label{fig:LGAD_hist}
    \end{subfigure}
    \caption{ (a) Signal charge pulses of the LGAD versus HV bias on the GaN photocathode under UVLED 4V, LGAD under 25V bias. (b) amplitude of the signal pulse versus the energy of the accelerated electron – a kinetic energy threshold of $\sim$1.5 keV was inferred.}
    \label{fig:design}
\end{figure*}

Throughout the experiment, we bias the LGAD at 25V operating in the linear charge gain regime and illuminate the photocathode with a constant UV average power of 110 pW. We collected  charge signals of LGAD at different accelerating electron energy by sweeping the high voltage bias on the GaN from 0V to 3kV, see Figure \ref{fig:LGAD_hist}.  
By plotting the amplitude of the signal pulse versus the energy of the accelerated photoelectron, see Figure \ref{fig:LGAD_hist},  a kinetic energy threshold of $\sim$1.5 keV was obtained.
Therefore, in this LGAD structure, it appears that $\sim$1.5 keV is the minimum kinetic energy required for the photoelectron to penetrate through the thin $\mathrm{SiO}_2$ passivation layer (or simply naturally oxidized) and generate electron-hole pairs in the silicon gain layer. For kinetic energies above this threshold up to the 3 $keV$ level, the response of the LGAD appears to be increasing linearly with the photocathode bias.

As shown in Figure \ref{fig:LGAD_hist} the large 40 mV LGAD signal was obtained with 3 $keV$ photoelectron, an LGAD bias of 25 V, and a UV LED drive pulse amplitude of 4 V at 100 Hz. 
Based on our initial calibration, the 40 mV amplitude signal is generated by an avalanche process which contains $\sim$34,720 electrons, achieving a signal-to-noise ratio of $\sim$9.5. 
While the QE of photocathode has slowly degraded to a value of about $6 \times 10^{-4}$, the average number of extracted electrons per UV light pulse was estimated at 818 and the average number of electrons reaching the LGAD at 18 per pulse. Therefore, the total charge gain for a single photoelectron is close to 2000.

\section{Conclusion and perspectives:}
In our study, we demonstrated the successful 
LGAD detection of photoelectron generated in GaN photocathode using UV photon. We constructed and characterized an HPD detector specifically designed for the detection of UV photons. The modular design allows for easy replacement of photocathodes and avalanche diodes. Our experimental results of the HPD provided valuable insights into the detection capabilities of the proposed GaN-LGAD setup under varying conditions. The initial measurement suggested the potential for large charge amplification in our HDPs, providing insight on venues for future improvement. 
To reach single keV electron detection level and to improve the signal-to-noise ratio, promising avenues include higher extraction voltage on the GaN photocathode, and lowering of the electronic detection noise.
By extrapolating the results from figure \ref{fig:LGAD_HV} to a larger  photocathode bias and specifically to a 5 kV voltage, we anticipate the LGAD signal amplitude 
would exceed 90 mV, and a single photoelectron could then produce a 5 mV signal, surpassing the current noise level of 4.2 mV. This approach presents a notable opportunity to further improve the signal-to-noise ratio.
Furthermore, placing a low noise charge amplifier circuit in vacuum, and in close proximity with the LGAD can lower the noise in the signal readout chain.
Increasing the QE of the photocathode by  improving in the in-situ cesiation process leading to NEA formation will increase single UV photon detection efficiency.

We believe all such improvements are within reach and if achieved could potentially pave the way to the deployment of a single photon HPD that can take advantage of already demonstrated highly efficient photocathodes and avalanche diodes capable of sub-ns time resolution, operation at hundreds of MHz, and low noise levels already at room temperature.

\acknowledgments

The authors gratefully acknowledge the support of the US Army (Phase-I SBIR contract W911QX23P0030), as well as the invaluable assistance provided by Mihee Ji and Anand Sampat at the US Army Research Lab. This work is also supported by BROOKHAVEN SCIENCE ASSOCIATES, LLC under contract DE-SC0012704 with the U.S. DOE.

\newpage


\bibliography{biblio.bib}







\end{document}